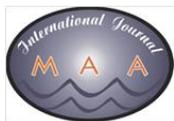 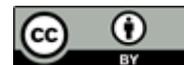

# RENUMBERING OF THE ANTIKYTHERA MECHANISM SAROS CELLS, RESULTING FROM THE SAROS SPIRAL MECHANICAL APOKATASTASIS

Aristeidis Voulgaris*[1], Christophoros Mouratidis[2], Andreas Vossinakis[3] and George Bokovos[4]

[1]Municipality of Thessaloniki, Directorate Culture and Tourism, Thessaloniki, GR-54625, Greece
[2]Hellenic Ministry of Education, Research and Religious Affairs, Kos, GR-85300, Greece
[3]Thessaloniki Astronomy Club, Thessaloniki, GR-54646, Greece
[4]Thessaloniki Science Center and Technology Museum-Planetarium, Thermi, GR-57001, Greece

## ABSTRACT

After studying the design geometry of the Antikythera Mechanism Saros spiral, new critical geometrical/mechanical characteristics of the Back plate design were detected. The geometrical characteristics related to the symmetry of the Antikythera Mechanism design, are independent to the present irregular deformation of the Mechanism parts and were used as calibration points for the Saros spiral cells positional measurements. The Saros cells numbering was recalculated using the calibration points position. A correction of minus one to the currently accepted numbering of the Saros cells was applied. Following the new numbering, a new proper position for the (displaced) Saros pointer axis-g, in graphic design environment was calculated. The measurements were tested on a bronze reconstruction of the Back plate, by the authors. This research leads to a new important result that the Saros does not start in a random or arbitrary date but only when a solar eclipse occurs within a month. Additional results were also calculated regarding the symmetry of the eclipse events/sequence. The new Saros cell numbering strongly affects the calculations for the initial starting date of the Saros spiral and the eclipse events scheme of the Antikythera Mechanism.

## 1. INTRODUCTION

The Antikythera Mechanism, a unique and remarkable geared machine of the Hellenistic era, was a time calculator and an astronomical event predictor of the solar and lunar eclipses. Today the Mechanism is partially preserved into six relative large fragments and 76 smaller pieces, totally corroded and deformed. An eclipse event information scale was located on the Mechanism's back plate: the Saros spiral, of four full turns and divided into 223 sectors/cells, representing the 223 synodic months of a Saros period. Some of the cells had engraved information relating to eclipse events - solar and lunar eclipses and the time they occurred (Freeth et al 2006; Freeth et al., 2008; Antikythera Mechanism Research Project, Anastasiou et al., 2016). The Saros pointer transited all of the cells after 241.029 full rotations (sidereal) of the Lunar Disk, which is the proper mechanical handling Input of the Mechanism (Voulgaris et al., 2018b; Roumeliotis 2018). Because of the strong and deep corrosion of the Mechanism's material, the broken Saros spiral is partially preserved in three parts (Fragments A, F and E), representing about 30% of the original part. The axis-*g* (and its gears) in which the (lost) Saros pointer was adapted, is also preserved on Fragment A (Anastasiou et al., 2014). Today, 22 eclipse information events are preserved on the Saros cells (Freeth et al., 2008; Anastasiou et al., 2016; Freeth, 2019; Iversen and Jones, 2019; Jones, 2020).

The Antikythera Mechanism is the result of celestial bodies' movement and time related phenomena observations, in a period hundreds of years before its Era. Astronomy regulated and directly affected the ancient Greek life (Hannah, 2015; Panou et al., 2020) and the religion (Liritzis and Vassiliou 2003; Liritzis and Castro 2013; Raul et al., 2018), and was very important for people around world (Freeth 2002a, b; Edmunds 2006).

Many references of eclipses around Mediterranean Sea and Middle East, by Haldean, Assyrian, Hittite, Egyptian and Greek astronomers, are preserved. In Ancient Greece territory, descriptions, recordings and information related to an eclipse are referred by *Homer* (Henriksson 2011; Papamarinopoulos et al., 2012), *Archilochus* poem (Lynn 1893), *Elikon the Kyzikean* (4th century BC), the philosopher *Thales*, *Conon the Samios*, *Hipparchus*, *Ptolemy* and *Plutarch* (Spandagos et al., 2000; Rovithis-Livaniou and Rovithis, 2007).

*Apollonious Rhodius* (Argonautica 1813, IV.58, p.275) and *Plato* 1967 (Gorgias 513a), refer that the famous *Witches of Thessaly* (around 400BC), were women that could predict the lunar eclipses (obviously using the Saros cycle), Dickie 2001. *Aglaoniki the Hegetor* (or *Aganice*, Αγλαονίκη η Ηγήτωρ, between 300-200BC), the daughter of the king *Hegemon of Thessaly*, could predict the lunar eclipses (Apollonious Rhodius 1813; Plutarch, Moralia, Coniugalia Praecepta, 48, p.340; Hill 1973). Even today, observation of eclipses still has a high scientific and social interest around the world for millions of people (Voulgaris et al., 2012; http://nicmosis.as.arizona.edu:8000/ECLIPSE_WEB /TSE2021/TSE2021WEB/EFLIGHT2021.html).

This work presents a new, precise calculation of the preserved Saros cells angular position of the Fragment A, leading to the renumbering of the cells with engraved eclipse events.

The method used is strictly based on the geometrical symmetry of the Mechanism design and it is not affected by the strong deformation of the parts, after 2000 years under the sea. The renumbering of the Saros cells leads to a number of new calculations, conclusions and results concerning the Saros eclipse events scheme and the initial Saros starting date.

The Greek word "*Ἀποκατάστασις*" (Apokatastasis, in singular) means return to the (same) starting position/place, as it was on its beginning, i.e. a reset position. It is extensively used on many sciences such as Medicine in Orthopedics Surgery e.g. apokatastasis of bones, in Archaeology e.g. apokatastasis of monuments-restoration and also in Astronomy. Ptolemy 1898 in Almagest refers the phrases "*Ἀποκαταστάσεις ἀνωμαλίας*", "*Ἀποκαταστάσεις πλάτους*", "*Ἀποκαταστατικός χρόνος*" (corresp. "returns in anomaly", "returns in latitude", Toomer, 1984 Book IV2, "time span between two successive resets to the starting position"). The word "*Apokatastasis*" is also referred many times on the Front cover inscription of the Antikythera Mechanism (Anastasiou et al., 2016b).

## 2. MATERIAL

### 2.1. *The Mechanical characteristics of the Saros spiral and its deformation*

The Antikythera Mechanism suffered irreversible chemical and positional changes during its 2000 years remaining underwater. The density of the bronze material (8.87g/cm$^3$) gradually changed to that of Atacamite (3.8g/cm$^3$), a product of copper/bronze corrosion under the sea water (Voulgaris et al., 2019b). The present day, the Antikythera Mechanism material is non-metallic, almost completely rocky (fossilized), even at its thickest parts (Fig. 1).

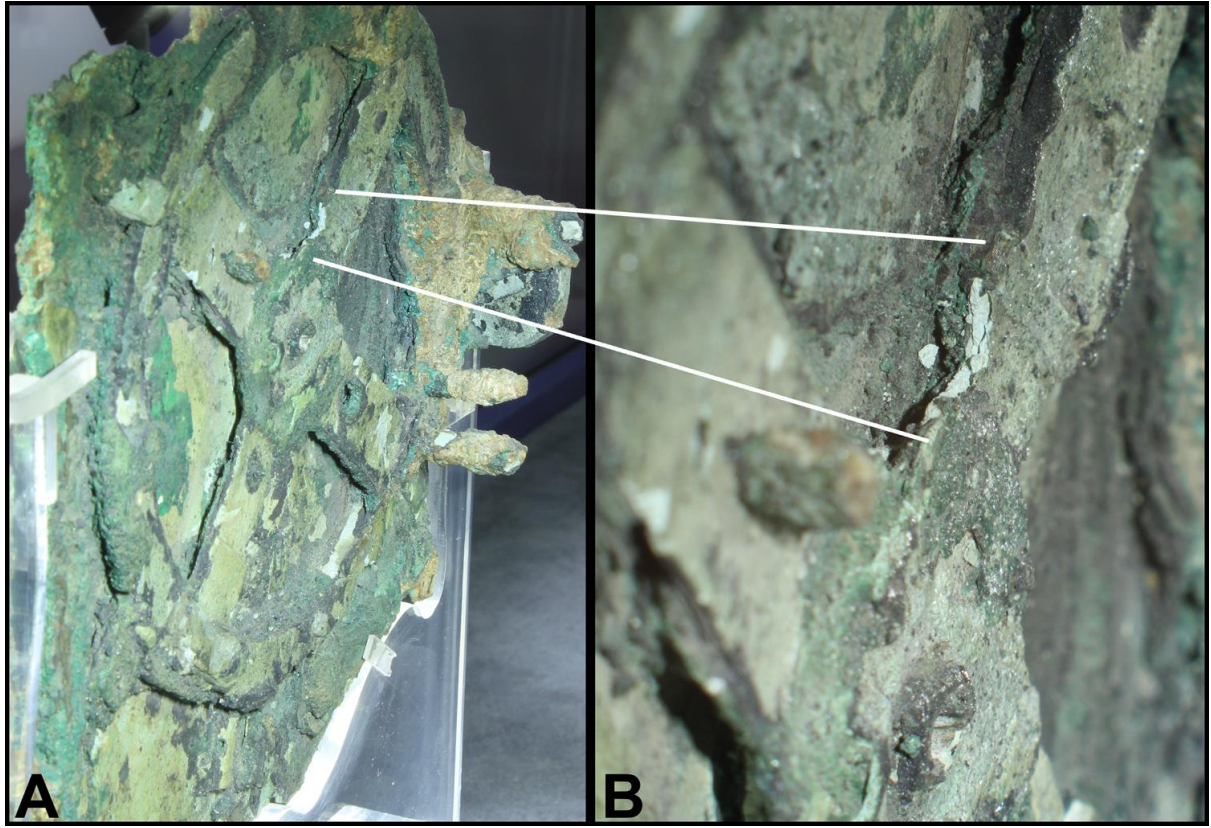

*Figure 1: A) An inclined side of view of the Fragment A1. The four arms of gear b1 are visible. B) On the 1 o'clock arm position of b1 gear, a random part is missing, revealing the internal cross section of the arm. The total absence of any bronze material is evident, also indicated by the degree of darkening of AMRP X-ray radiographies and tomographies (Voulgaris et al., 2018c). The deep inside corrosion product of Atacamite (instead of bronze) is today the main material of the Antikythera Mechanism fragments (Voulgaris et al., 2019b). Credits: National Archaeological Museum, Athens, Copyright Hellenic Ministry of Culture & Sports/Archaeological Receipts Fund Photos. Images by the first Author.*

After the sudden retraction of the Mechanism from the sea bottom and its contact with the dry atmosphere and the wet environment, the Mechanism (today in fragments) abruptly and irregularly shrunk, leading to cracking, breaking and fragmenting. The shrunk fragments were distorted, displaced, deformed in 3D direction, the straight sides are "wavy" like, the plates are no longer level and the present position of the parts differs than their original (Voulgaris et al., 2019b). It is obvious that the present dimension of the Mechanism's parts (gears, plates, rings etc.), are smaller than the original as a result of the corrosion and shrinking.

Although the deformation of the Mechanism is irreversible and some parts are totally missing, the remained mechanical evidence highlights the instrument's design and its Symmetry, which can be useful especially for the dimensional measuring, testing and the positional Apokatastasis of some parts.

Saros and Metonic spirals are located on the Back plate, but today they are both partially preserved (Anastasiou et al., 2016b). The Saros spiral consists by four-full turns divided in 223 subdivisions/cells. The spiral turns, cannot be stable and fixed on their position without an additional support component (Fig. 2A). In order to immobilize the spiral turns on their position, the ancient manufacturer adapted three retention bars, positioned in epicenter angles of 120°, measured CW and centered on spiral axis/pointer (axis-$g$ for Saros spiral and axis-$n$ for Metonic). For Saros spiral, the distribution of the retention bars is -60°, +60°, +180° relative to the vertical *VLb* line and -120°, 0°, +120° for Metonic spiral (see next Chapter), as can be calculated by the AMRP tomographies (also in Wright, 2005a; Wright, 2005b; Voulgaris et al., 2019b).

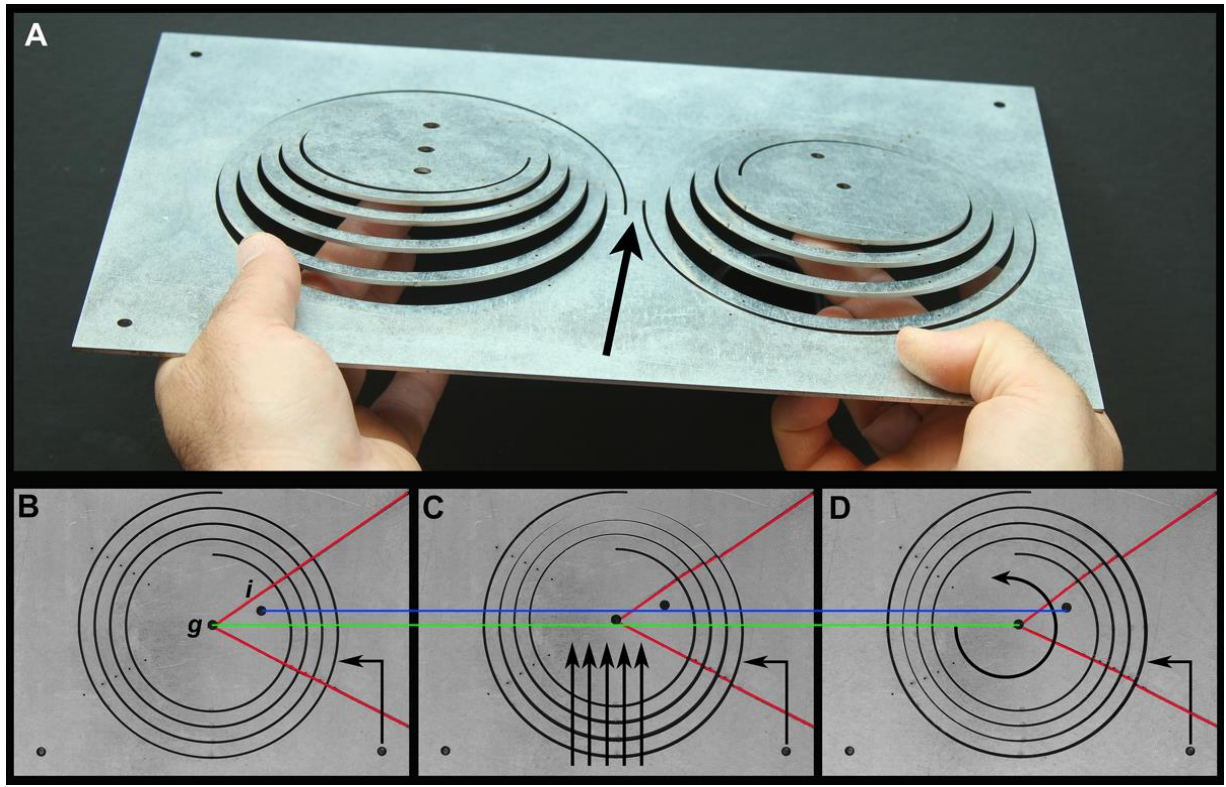

*Figure 2. A) An authors' reconstruction of the Back plate was used for the deformation and displacement study of the spiral parts. To study the mechanical behavior of the Metonic and Saros spirals, the retention bars were subtracted (in this case eight bars in total for measuring reasons). By applying a pressure perpendicular to the plate, the spirals are also deformed. The mechanical stability of the Back plate will be additionally decreased, if the ends of both spirals are connected. B) The spiral turns are located on the proper position and two lines by a permanent black pencil, were marked on the Saros spiral (on the photographs, a red line is superimposed on the black lines for visibility reasons). C and D) By applying a pressure in the direction showed by the black arrows, the 2nd and 3rd spiral turns are displaced. Also the holes for the axis-g (Saros pointer) and i (Exeligmos pointer), after the pressure, changed their position. Note that the position of the 4th spiral turn, as a part of the massive material of the Back plate is not displaced (see the position of the black arrow relative to the low right plate hole). The shafts' (f), g, (h), i position relative to the rest massive Back plate can be varied after applying a pressure, but their between relative position/axial distance, is more difficult to be strongly affected.*

The retention bars are stabilized on the Back plate using secret pins-without head (Fig. 3C and Fig. 6). The profile shape of the retention bar was designed as a "*sinoid*" or a "*square pulse graph*" and constructed in this way in order not to disturb/block of the Saros pointer edge, which travels inside the spiral rim. The existence of the spiral rim reduces the mechanical stability of the spiral(s), which do not behave as a solid bronze material. On the contrary, the area outside the spirals (4th spiral turn of Saros dial and 5th spiral turn of Metonic dial), are fixed, as parts of the rest massive Back plate (Fig. 2B,C,D).

The poor mechanical stability of the Saros spiral resulted to its deformation and displacement. As is visible on the AMRP tomographies and the 3D reconstructions by the authors, the Saros right hand spiral turns, which are located on Fragment A are distorted, broken and displaced. In addition, their geometrical centers are not perfectly coincident and they are not located on the expected position. Moreover, the spiral rims have different widths. The Metonic spiral part (Fragment B), is preserved in better condition (Anastasiou, 2014; Allen et al., 2016). The Saros spiral turns are not located on the same level and the preserved retention bar is also distorted and broken, as is evident on the 3D reconstructions (Fig. 3A,C and Fig. 4). Furthermore, the central area of Saros spiral with the gears/shafts *f*, *g*, *h*, *i*, is not parallel to the (also deformed) Middle plate plane (Fig. 4).

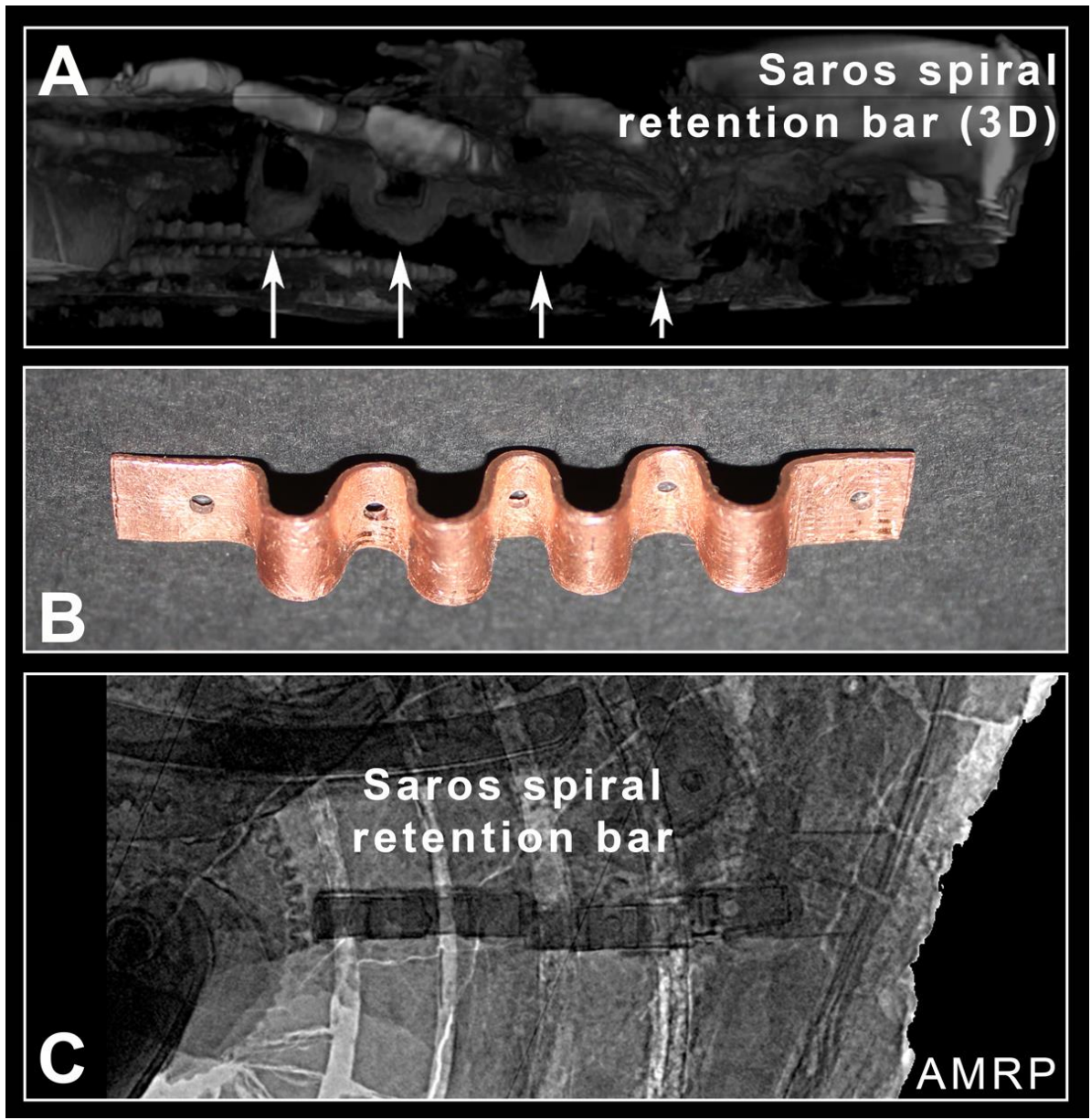

*Figure 3. A) 3D visualization of the Saros spiral retention bar using the AMRP tomographies and the software 3D Slicer (Fedorov et al., 2012), processed images by the authors. B) Reconstruction of a Saros spiral retention bar. The "square pulse graph" shape of bar is visible. C) The distortion of the Saros spiral retention bar, is evident. AMRP positive radiography processed by the authors.*

As results from the tomographies and the 3D reconstructions (Fig. 4A,B), the Saros axis-$g$ is not perpendicular to the Middle and Back plates, so the Saros axis-$g$ present position differs from the original, as well as the spiral turns. Therefore, it is too doubtful to consider that the Saros axis-$g$ has remained fixed on the same original position after the Mechanism deformation and shrinkage.

The non-uniform distortion, the irregular deformation and displacement of the parts, make it difficult to measure and calculate the original dimensions and position of the Mechanism parts because it is difficult (or impossible) to be found non-displaced/non-deformed points, so that to be considered as the fixed/dimensional calibration points. In order to improve the estimated dimension/position of the parts, their mechanical characteristics, limitations and requirements must be taken into account, applying dimensional and positional corrections so that the reconstructions of these parts work properly. At the same time the geometrical mechanical characteristics of the original artifact, must be taken into account, considering the geometry of the design Symmetry, which was extensively used by the ancient Greeks see Chapter 4.

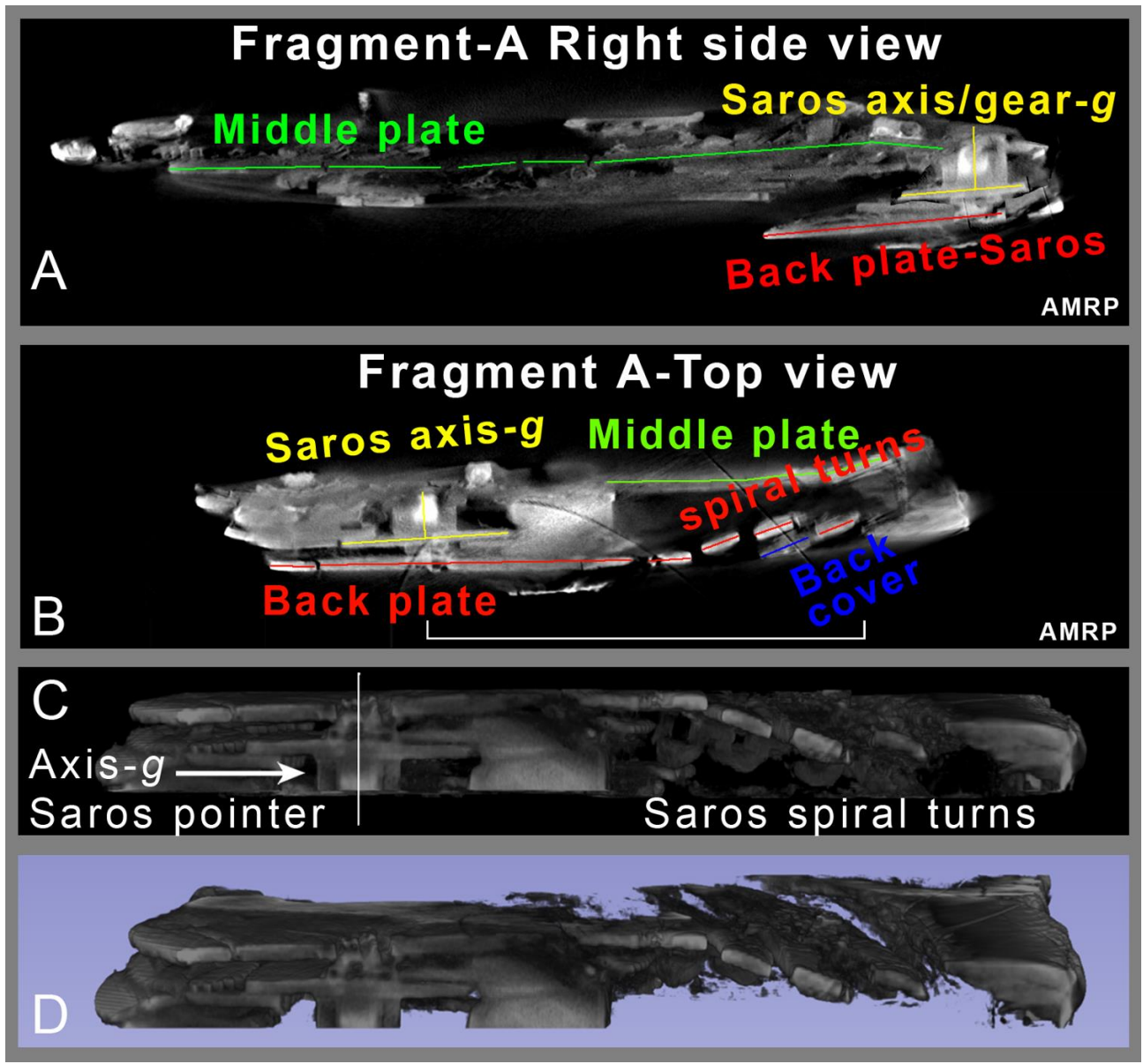

*Figure 4: A) On a right side view tomography of the Saros spiral central area, the gear/axis-g was digitally added on its corresponding position. The Middle plate cross section is sketched in green color line with gaps (where the plate is broken), the Back plate preserved Saros sketched in red line and the Saros axis/gear-g sketched in yellow line. The positional deviation of the parts from their original "straight/perpendicular" position is evident. B) A single, top-view tomography on the Saros axis-g area. As is evident, the Saros spiral right hand turns, are strongly deformed, in about 20° inclination than the "ideal" plane defined by the Saros central area. The outer limits of the Back and Middle plates are also deformed. C and D) In the 3D visualization of the Saros spiral turns area, using the AMRP tomographies and the software 3D Slicer (Fedorov et al., 2012), the extended deformation of the spiral turns, is evident (C: cross-section, D: by a small front inclination). AMRP tomographies processed by the authors.*

## 3. EQUIPMENT

Measurements and calculations were based on the Antikythera Mechanism visual photographs, AMRP PTM and ours. At first, the visual photographs were calibrated and afterwards correlated to the AMRP Computed Tomographies (CTs), on the areas of interest. For the dimensional calibration of the material, photos of the AM by Xenikakis (2005) and by the authors were used. Camera lenses aberrations, such as field curvature, barrel or pincushion distortion or parallax projection - i.e. the specific position of the optical axis of the system camera/lens relative to the center of the fragment(s) and also the angle between the optical axis and the objective's plane (ideal at 90°) (Hecht 2015), can lead to optical defects. In order to suppress all these optical defects, care was taken to matching only specific areas of interest to the corresponding CTs and not to the full area of the large Fragment A.

Several design software were used for the dimensional linear measurements on the calibrated photographs and CTs. The final measurements were achieved by three different visual images of the same area, which gave similar results. Using the same procedure on the oldest photographs by A. Rehm 1906[1] (significant geometrical distortions were detected, as a result of the positional parallax effect (and also by the old design camera lens that was used).

Additionally, the 3D digital reconstructions, using AMRP CTs, with the 3D Slicer reconstruction software (Fedorov et al., 2012), was a significant helping aid to calculate the degree the deformation/displacement in 3D space.

Our final measurements and the Saros axis repositioning, were calculated by a computer-aided design software (CAD) for architects, engineers and construction, capable to create precise 2D and 3D drawings.

## 4. MEASUREMENTS AND RESULTS

### *4.1 Two critical observations for the Saros spiral position*

#### 4.1.1 Observation A

Symmetrical designs and structures are common phenomena in natural and artificial constructions (Landau et al., 1967; Stewart, 2013; Zhiyong et al., 2015). The ancient Greeks used extensively the idea of the design Symmetry on their architecture and constructions, applying the Euclidean Geometry (Vitruvius, 1914, Book III; Coldstream, 1991; Orlandos, 1994; Lloyd, 2010). By following this principal, they made complex designs and constructions based on the Geometry (Duvernoy, 2018). Symmetry also offers an aesthetically beautiful balanced result and art work satisfaction (Osborne, 1986; Hambidge, 1967).

The Antikythera Mechanism design illustrates the principle of the Geometry and the Symmetry. For example, consider the axis-$b_{in}$ located on the central area of the Front plate, the Lunar Disc, the annual gear *b1* and the Golden sphere-Sun rotated around this center (Voulgaris et al., 2018b). The two centers of the Egyptian and Zodiac month dial scale rings coincided with the axis-$b_{in}$. The two Parapegma plates are symmetrical located on top and bottom of the Front Dial plate (Bitsakis and Jones, 2016b). The two Back plate spirals also exhibit geometrical symmetry. For this reason, the ancient manufacturer designed them in approximately same outer dimension (Allen et al., 2016), and the spiral centers crossed by the median Vertical Line-*VLb*. The Symmetrical design of the exterior face of the Mechanism defines the specific position of the axes/holes inside the Mechanism.

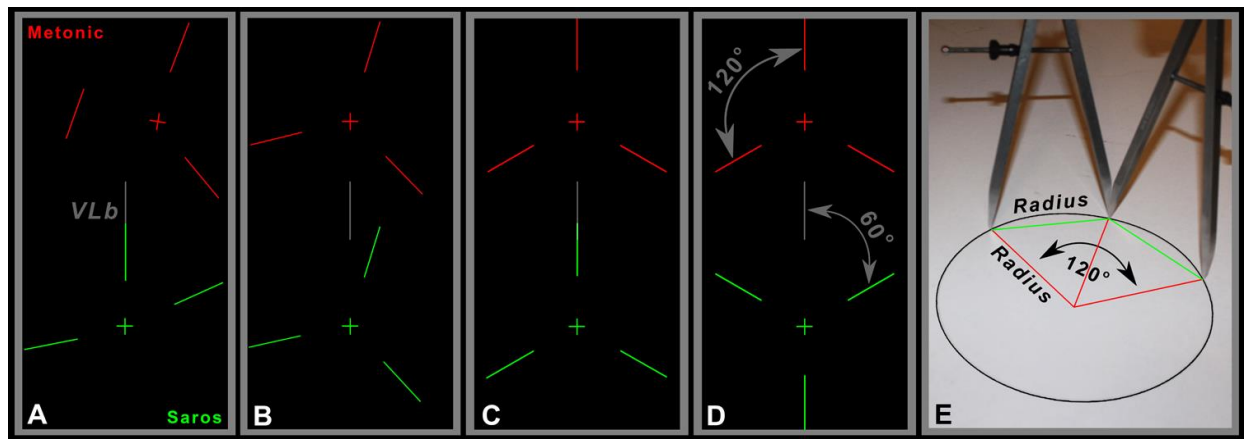

*Figure 5. The distribution probability of the six retention bars on the Back plate. In red, the distribution of the Metonic spiral retention bar and in green the retention bars of Saros spiral: A) A non-symmetrical, random distribution. B) Assuming a random symmetrical distribution (Translational Symmetry), but without symmetry relative to the VLb axis (grey line). C) A symmetrical distribution of the retention bars, relative to the VLb axis. D) Distribution in balanced symmetry (axial and reflection) relative to the VLb and the horizontal axis. By the study of the preserved fragments of the Mechanism, the ancient manufacturer used the distribution of the balanced symmetry (D). E) Designing an epicenter angle of 60° and 120°, using a simple compass (image by the authors).*

The Saros (and the Metonic) spiral turns are stabilized by the use of 3 retention bars. The position of the preserved Fragment A2 retention bar and the absence of a bar on Fragment F, considering the geometrical symmetry: the ancient manufacturer placed the three retention bars spaced by 120° around the spiral and at the same time the one out of three retention bars is coincided to the *VLb* line (see Fig. 11 of Wright, 2005a, Fig. 7 of Wright, 2005b, and Fig. 7 of Voulgaris et al., 2019b). This presents a balanced Symmetry, (Fig. 5).

---

[1] A. Rehm, Untitled manuscript dated 1906 ("Athener Vortrag"), Bayerische Staatsbibliothek (Munich), Rehmiana III/9.

Hence, it turns out that the preserved Saros spiral retention bar on Fragment A2, should be located at an epicenter angle of 60° CW relative to the vertical line *VLb*, which crosses the Saros axis-*g*, (Fig. 5D). All of the Saros retention bars were stabilized on the Back plate by five secret pins for Saros spiral (and six for Metonic), which should be also placed at an epicenter angle of 60° CW relative to the vertical line *VLb* (Fig. 6).

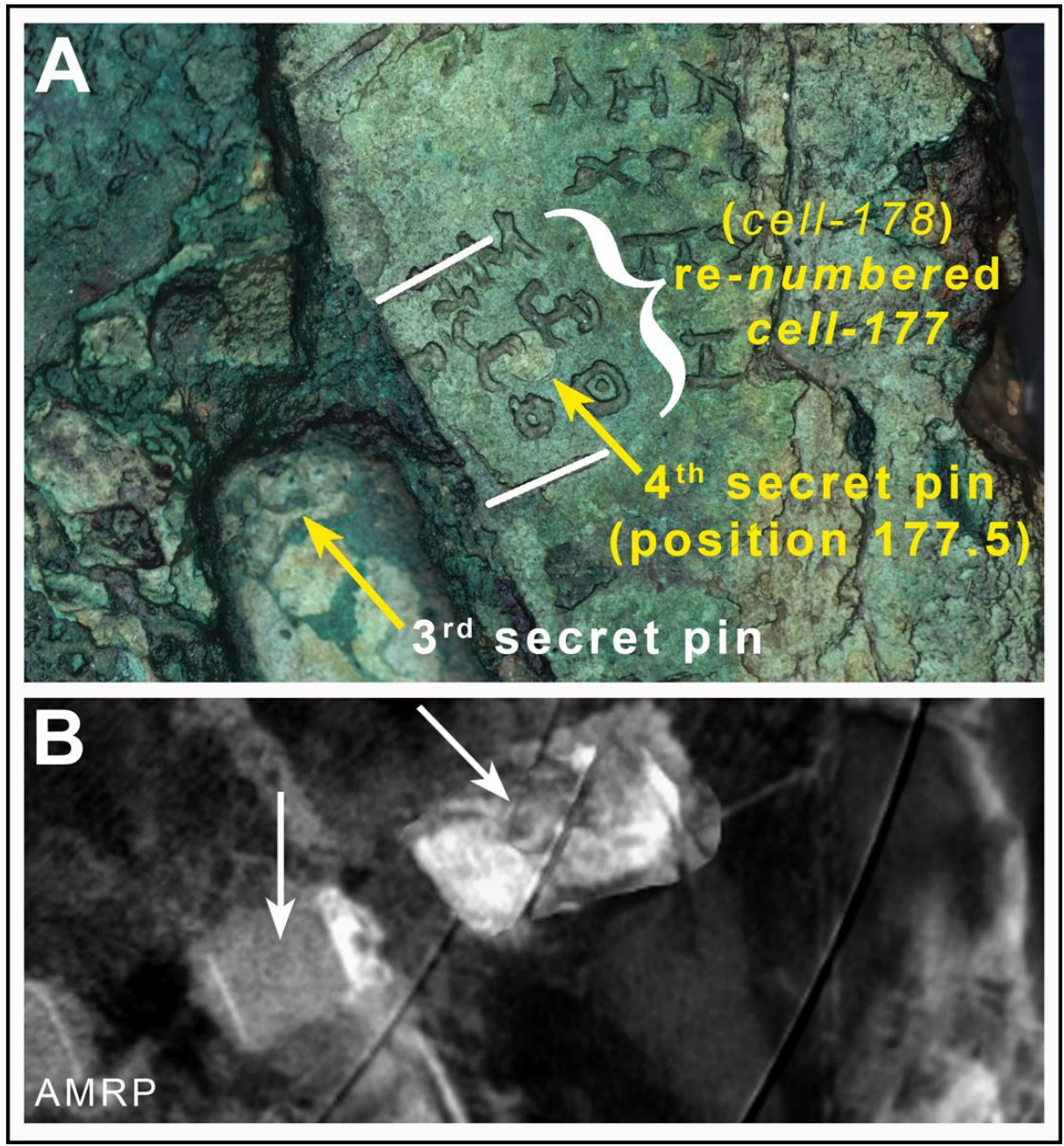

*Figure 6. A) AMRP PTM close-up photo of cell-178, renumbered cell-177, area. The secret pin-4 of the retention bar, in the middle of the cell, between the Solar and Lunar eclipse event information, is clearly visible on the Back plate. A part of the "hour glyph" is engraved on the secret pin and therefore the ancient manufacturer firstly stabilized the spirals via retention bars and then engraved the eclipse events. The secret pin-3 is also visible on the 3rd spiral. B) the corresponding combined AMRP tomography at the same scale, depicts the secret pins on the retention bar.*

The characteristic geometry of the angle 60°/120° is preferred by engineers, because it can be very easily designed: dividing a circle in 3X120° can be accurately achieved as 2X60°(X3) using a simple compass (a cord of an arc, equal to the radius offers an epicenter angle of 60°, (Fig. 5E). Note that the ancient manufacturer used a compass during the construction of the Mechanism (see Fig. 6 of Voulgaris et al., 2019a).

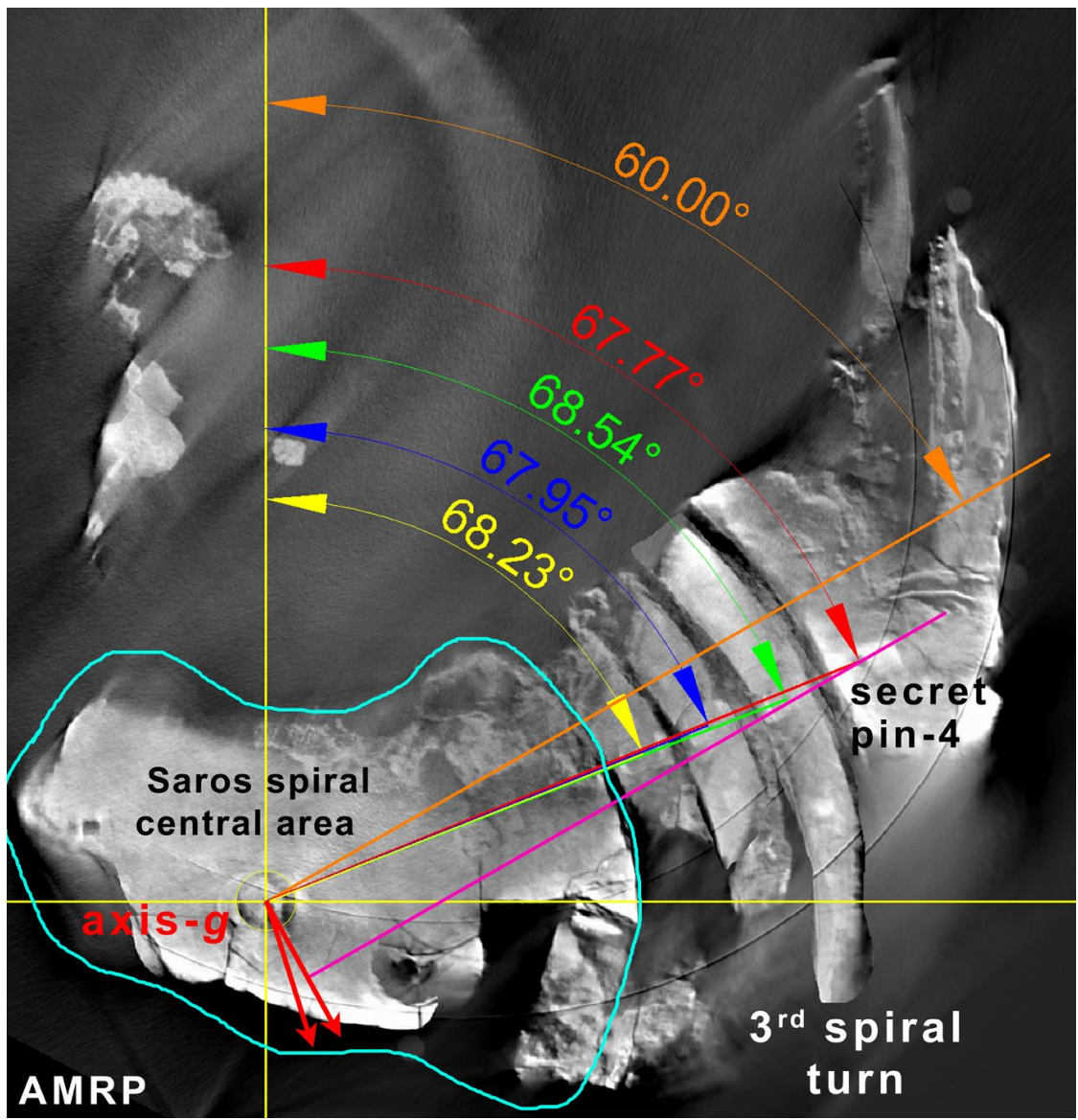

*Figure 7. AMRP multi-combined tomography. Axis-g and axis-$b_{in}$ crossed by VLb-line. The calculated epicenter angles on the present (deformed) position of axis-g and the secret pins (in red, green, blue and yellow color), relative to the calculated ideal and proper epicenter angle of 60° (orange color). The present deviation of the epicenter angle is about 8° from the ideal value of 60°. By shifting the Saros spiral central area (delimited by the turquoise line) and therefore the axis-g to the direction of the two red arrows (the one is perpendicular to the orange line), the epicenter angle decreases and approaches the angle of 60° (purple line). Image processed by the authors.*

From Fig. 7 results that the present day epicenter angle of the five secret pins, measured on the original artifact, differs from the expected 60° angle relative to *VLb*, as a result of the Saros spiral deformation and the Saros axis displacement. Assuming that the ancient manufacturer made a measuring error on the positioning of the pins seems improbable, as the difference 67.77°−60°= 7.77° corresponds to a difference of about 10 mm on the perimeter that crosses the last pin-4, which is quite large.

By observing the predefined symmetrical design of the positions of the retention bar and the secret pins, located at an epicenter angle of 60°, a very precise calculation of the cells index number can be achieved.

This calculation is not affected by the (displaced) position of axis-*g* or by the distorted position of the spiral turns.

The numbering of the preserved Saros cells can be calculated by the corresponding spiral turn, the epicenter angle φ between the vertical line *VLb* which crosses the axes $b_{in}$ and *g* and the line connecting the center of axis-*g* to the corresponding Saros cell-first subdivision line, by the following equation (1):

cell-y fraction = [1+(55.75 X number of full spiral turns) + φ/6.45739]     (1)

(as cell-1 is defined the first boundary line of the 1st cell, which is also crossed by the *VLb* line).

*Table 1. The calculated cell fraction number of the four secret pins, based on their epicenter angles (relative to VLb line), measured on the present condition of Fragment A2 and the present position of axis-g (initial cell numbering published by Freeth et al., 2008; Carman and Evans, 2014; Anastasiou et al., 2016; Freeth, 2019; Iversen and Jones, 2019; Jones, 2020). The next column lists the same calculation by setting the secret pin epicenter at a fixed angle of 60° relative to the VLb line. The last column lists the correction factor on the initial cell numbering according to the observations of the present work.*

| Secret pins on the retention bar | Epicenter angle of secret pin measured on Saros spiral present condition (deformed) (Fragment A2) → Calculated cell fraction according to equation (1) → (corresponding cell) | Applying epicenter angle 60°→ Calculated cell fraction→ (corresponding cell) | Correction factor in initial cell numbering after the Saros Apokatastasis |
|---|---|---|---|
| Pin-1, (1st turn) | 67.77°→ cell fraction-11.49→ (cell-11) | 60°→cell fraction-10.29→ **(cell-10)** | **-1** |
| Pin-2, (2nd turn) | 68.54°→cell fraction-67.36→ (cell-67) | 60°→cell fraction-66.04→ **(cell-66)** | **-1** |
| Pin-3, (3rd turn) | 67.95° →cell fraction-123.02/122.99 → (cell-123/122) corrected to cell-122 | 60°→cell fraction-121.79 → **(cell-121)** | **-1** |
| Pin-4, (4th turn) | 68.23°→cell fraction-178.81→ (cell-178) | 60°→cell fraction-177.54 → **(cell-177)** | **-1** |

By defining the fixed epicenter angle φ=60°, the corresponding cell fraction in which a secret pin is located, is calculated according to equation (2):

cell-y fraction = [10.29+(55.75 X number of full spiral turns)]     (2)

In this way, the cell numbering calculation is very precise, as the specific position of a secret pin can be measured to sub-cell accuracy/cell-fraction. E.g. equation (1) yields a value of 177.5 for the secret pin on the 4th spiral, a value that matches perfectly to the original pin position, visible directly by naked eye on Fragment A2, (Fig. 6 and Fig. 8). In contrast, by applying the present epicenter angle for the pin-4 (Table 1) yields a cell number of 178.81, which is not in accordance to the present pin position, (Fig. 8).

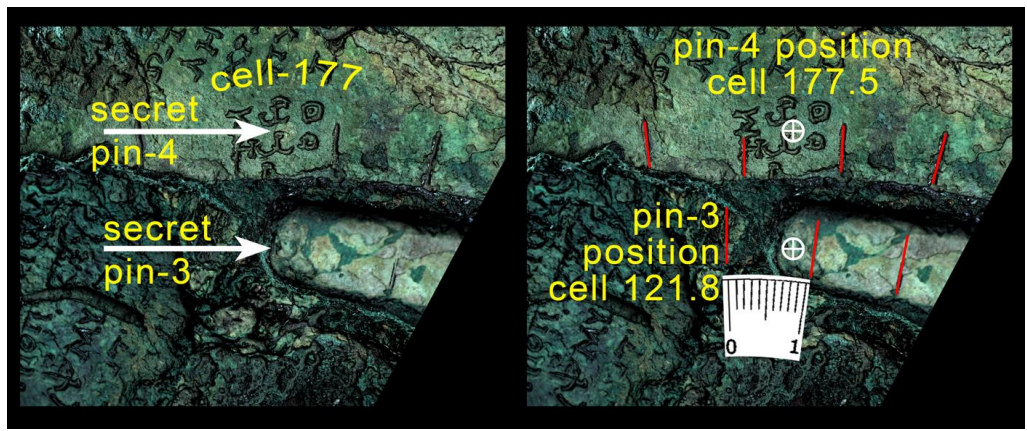

*Figure 8. AMRP PTM image. A close-up of the 4th and 3rd Saros spiral turn. The secret pins 4 and 3 for the stabilization of the retention bar, are visible inside the cells 177 and 121 respectively (new cell numbering), considering the secret pin positions in epicenter angle of 60°. The corresponding cell fractions are 177.5 and 121.8.*

Table 1 lists the calculated secret pins position on the corresponding cell fraction (eq. 2), compared to the real pins position, calculated either by naked eye observation or detected/calculated on the AMRP topographies. The results of the measurements for the secret pin(s) positions match very well to the theoretical results based on the AM symmetrical design. The present work observations and calculations lead to a correction on the initial numbering of the Saros cells, published by Freeth et al., 2008; Carman and Evans, 2014; Anastasiou et al., 2016; Iversen and Jones, 2019; Freeth, 2019; Jones, 2020.

*Table 2. The numbering of the cells in which there is a secret pin, setting the epicenter angle to 60° relative to the VLb line. The calculation concerns four cells corresponding to the spiral turns. The secret pins cell fraction position at an epicenter angle of 60° matches almost perfectly to the original position of the pins.*

| Spiral turn | Secret pin position on cell fraction, calculated in epicenter angle 60° | Observed/measured or calculated position of the secret pin on cell fraction (photographs/AMRP CTs) | Corresponding integer cell number |
|---|---|---|---|
| 4th | (177)+0.54 | ≈cell+0.5 (photograph) | 177 |
| 3rd | (121)+0.79 | ≈cell+0.8 (photograph) | 121 |
| 2nd | (66)+0.04 | ≈cell+0 (interpolated by the boundaries of cells 177, 121 and 10 (CT) | 66 |
| 1st | (10)+0.29 | ≈cell+0.4 (calculated on the tomography: ≈1.6 cells before the 1st boundary line of cell-12 | 10 |

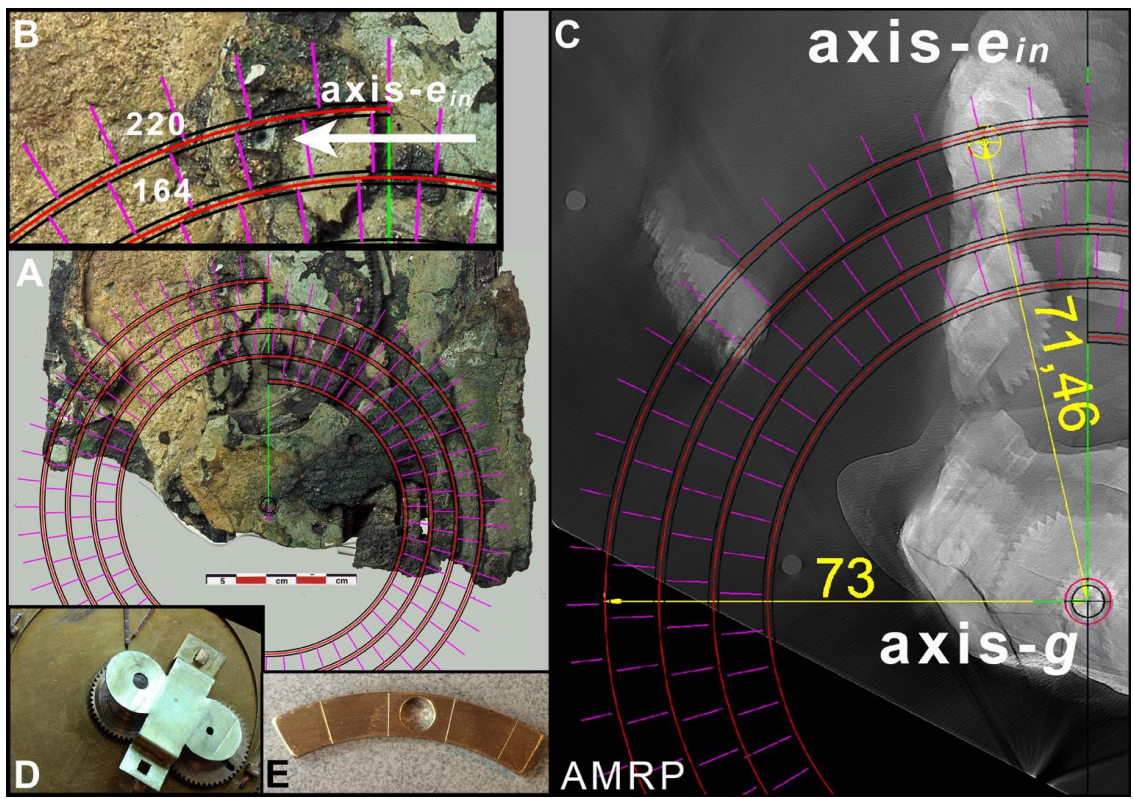

*Figure 9. A) same scale composite image of the bottom area of Fragment A2 (Xenikakis, 2004) and a sketch of the Saros spiral by the authors (in black color lines are the rim boundaries, in red the rim middle boundary distance and the green line is the VLb vertical line). B) Close-up of the image. The axis-ein is located on the cell-166 (considering the present position of Saros spiral), close to the 4th left hand rim. Photograph credits: National Archaeological Museum, Athens, K. Xenikakis, Copyright © Hellenic Ministry of Culture & Sports/Archaeological Receipts Fund. C) The Saros spiral designed by the authors added on the AMRP multi-combined tomography processed by the author(s). The radius dimension of the 4th left hand larger spiral turn and the axial distance g-ein and the distance of axis-g by are presented in yellow color. The axis-ein is also located on the cell-166 area. D) the Ω-shape retention bar before its adaptation on the e3 gear. Bronze reconstruction and image by the author(s). E) a representation of a hypothetical 5 mm hole on the middle of the spiral width, for the axis-ein. The hole covers the large percentage of the cell area and any engraved information is impossible.*

### 4.1.2 Observation B: Confirming the Observation A

Considering the present position of the Saros axis-$g$, the author(s) designed the Saros spiral using the calculated dimensions by Allen et al., 2016 (also Anastasiou et al., 2014; Freeth et al., 2008; Freeth, 2019). Then, several designs of the Saros spiral were placed to the (same scale/aligned) visual photographs (Xenikakis, 2004; AMRP PTM; author, 2013-2019) and AMRP tomographies.

During the study, it resulted that the position of the axis-$e_{in}$, is located just on the 3rd left hand spiral turn, between cell(s)-166/(165) or just on the 4th left hand rim of the Saros spiral or between the rim and the cell-166, (Fig. 9). The axial distance $g$-$e_{in}$ was measured by the author(s) between 71.4 mm-73 mm and the radius of the largest slot of the Saros dial is about 73 mm (Allen et al., 2016).

The long length axis-$e_{in}$ is a very important axis of the Mechanism, because many gears and one additional axis are adapted on it: the gears $e1$, $e6$ as well as the axis-$e_{out}$, with gears $e2$, $e3$ and $e5$, are supported on axis-$e_{in}$ (on the gear $e3$, gears $e4$, $k1/k2$ are also supported, Freeth et al., 2006). Most of the Mechanism axes are supported on the Middle and the Back plate opposite holes (except the self-supported on the Middle plate, small diameter gears $c1,2$ and $l1,2$ which have very short length axis). The holes on the Back plate can be easily detected on the visual photographs (on Fragment A2, holes for axes $f$, $g$, $h$ (Voulgaris et al., 2018b) and on Fragment B for axis $n$) and on the AMRP tomographies (for axes $i$ and $o$).

The supporting of long-length gear axes/shafts on two bearing points, offers a high stability avoiding gear disengagement and it is a common design for the gear reducers, even today (Fig. 10). The axis-$e_{in}$ is quite large and therefore must be supported on two points i.e. on the Middle (as is) and on the Back plate. Otherwise, the rotation of the axis and its gears is difficult or even impossible.

Another, but not very satisfactory way is the stabilization of the axis-$e_{in}$ in one point (on the Middle plate as is) and the other edge to be (self) stabilized by a perpendicular pin, just after the e6 gear and before the Ω-shape retention bar. However, this cancels the reason of the Ω-retention bar existence, which stabilizes the gears $e5$, $e6$ (or just after the Ω-shape retention bar, which also cancels the reason of the bar adaptation). The ancient manufacturer adapted the Ω-shape retention bar on the gear e3 in order to maintain gears $e5$-$e6$ and $k1$-$k2$ on their positions (see Fig. 9D and also Fig. 12 of Wright, 2005b).

Today only one base of this Ω-shape retention bar is preserved (Voulgaris et al., 2019a). After the careful examination of the AMRP tomographies and the visual photographs, there was no evidence of the existence of any perpendicular stabilizing pin (right before/after the Ω-shape retention bar) or a perpendicular hole at the well preserved, edge of the axis-$e_{in}$ and therefore these two solutions do not seem possible.

This long length axis is also rotating and it is not fixed on the Middle plate, therefore its support on two edges is necessary. Moreover, by supporting it on two points, the gears $e2$, $e3$, $e5$ and $e6$ remain steady on their positions and the axis-$e_{in}$ remains perpendicular to the Middle plate, during its rotation.

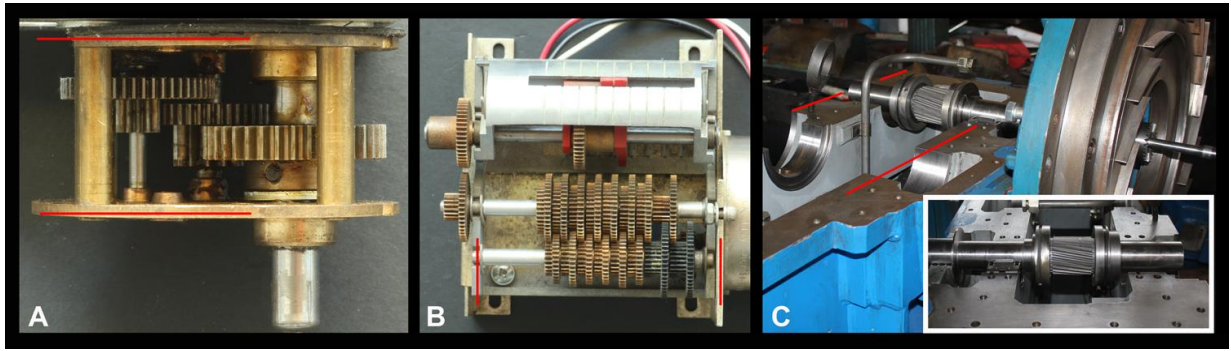

*Figure 10. Typical designs of geared reducers. The red lines depict the two (necessary) parallel plates in order to support the two edges of the gear axes. A) A simple geared reducer. Its axis output is also visible. B) A multi-variable reducer. C) The opened box of an air turbine. The output gear/axis is supported on two (opposite) points. On insert, close up of the gear and its supporting areas. Gear reducers and images, by the first author's collection.*

During the reconstruction of the Antikythera Mechanism Back plate by the author(s), considering that the Saros axis-$g$ is located on its present position, an attempt for the axis-$e_{in}$ edge support of on the Back plate, by drilling a hole in diameter ≥5 mm (the axis edge has square cross section in diagonal 5mm) just on the calculated position, was made: either by drilling a hole just on the 4th spiral rim, either by drilling the cell-166/(165) area.

Just a hole on the rim, cannot be realistic.

By drilling just on the cell-166 central area, a serious mechanical problem arises: the bronze material width each of the Saros spiral turns is about 6.7 mm (Allen et al., 2016). So, 6.7 mm (spiral width) – 5 mm (hole)= 1.7 mm, i.e. the solid material up to the rim boundaries is only ≤0.85 mm (Fig. 9). This hole, will additionally downgrade the poor mechanical stability of the spiral turns and this unstable behavior having an impact on the axis-$e_{in}$ stabilization. Moreover, on the cell-166 has been predicted a solar eclipse event (index event Ξ/*symbol*, see Table 4.6-Revised glyph sequence in Anastasiou et al., 2016, also Freeth, 2019; Iversen and Jones, 2019) and if there was a hole, the event engraving is impossible (Fig. 9E).

The only satisfied, logical and mechanical accepted solution on this serious mechanical problem, is that the ancient manufacturer avoided drilling any of the spiral turns and he selected a position for drilling a hole for axis-$e_{in}$, in an area of the Back plate with solid mass of bronze, out of the three first - left hand - spiral turns and far away from the outer limit of 4th left hand rim.

If the Saros axis-$g$ was on its present position, the ancient manufacturer could easily avoid the crossing of axis-$e_{in}$ with a Saros spiral turn or cell, by changing the $e_{in}$-axis position and its corresponding hole on the Middle plate, in a higher location, without any problem on the gear engagement. As a result of the corrosion, deformation, the long stand still on the sea bottom, the effects of gravity and the shrinkage of the Mechanism, leads to the conclusion that the present position of the Saros spiral is not its original position (see also the deformation of the front central area of Fragment A in Fig. 6, Voulgaris et al., 2019b). Therefore, the Saros spiral turns and the Saros axis-$g$ should be repositioned in a different place than the present.

The author(s), using the calibrated radiographies/images and graphic design software, researched new proper positions of the Saros axis-$g$, applying the minimum possible positional changes of the parts, in order the adaptation of the axis-$e_{in}$ edge on a hole of the Back plate, to be mechanical accepted, realistic and functional on the model reconstructions. An estimated positional correction of the displaced axis-$g$ and the Saros spiral central area is presented on Fig. 7 (see the direction of the red arrows, 20°-30° CCW by the *VLb* line). By repositioning of Saros axis-$g$, in about 6mm, the hole for axis-$e_{in}$ located inside *cell-221/222* and the epicenter angle on the pins is about 63.4°.

A hole just inside a cell (even without an eclipse event), which is a subdivision of an important measuring scale, does not seem to be an ideal position, because it decreases the quality and the aesthetic of the outer design of the Mechanism (Fig. 9E, this is equal to a large stabilizing screw of a car dashboard visible just on the *km/h* subdivisions of the car tachymeter).

The hole for the axis-$e_{in}$ is totally out of the Saros boundary cells, if the repositioning of Saros axis-$g$ is ≥9.8 mm by its present position, the epicenter angle of the pins is around to 60.4° and the author(s) believe that it is the proper and close to the original position of the Saros axis-$g$.

The Observation B, confirms the observation A.

The epicenter angle measurement can also be used in reversed manner: what the relocation of the Saros axis-$g$ should be, so that the retention bar pins to be located on the epicenter angle, 60° relative to the *VLb* line (Fig. 7).

By repositioning the Saros spiral by 9.8 mm to the direction of the red arrows (projected repositioning in *VLb* line about 8.5mm), the total distance between the Metonic pointer axis-$n$ and Saros pointer axis-$g$ is 145.5 mm (Allen et al., 2016)+8.5 mm (projected reposition)= 154 mm. The central radius of the largest slot of the two spirals is about equal of 73 mm (Allen et al., 2016). Therefore, 154 mm−(2X73)= 8 mm (the minimum distance of the two spiral rims). From this argument is resulted that the two spiral rims do not intersect and the Metonic and Saros spirals are independent, giving at least a relative stiffness of the Back plate, see Fig. 3 and on the same time the stirring of the Metonic and Saros cells is avoided (see Fig. 6 of Wright, 2005a).

## 5. DISCUSSION

### 5.1 *The New cell numbering after the Saros axis positional Apokatastasis and the Sar period*

The observations and the results of Chapter 4, alter the initial Saros cell-numbering and the eclipse information scheme, presented elsewhere (Freeth et al., 2008; Freeth, 2014; Anastasiou et al., 2016b; Freeth, 2019; Iversen and Jones, 2019 and Jones, 2020). The new cell numbering of the Saros events, Fig. 11 also challenges the calculation of the Antikythera Mechanism Epoch Date discussed by Carman and Evans, (2014); Freeth, (2014); Iversen and Jones, (2019); Jones, (2020).

Table 3 presents the recalculated cell numbers for the cells that have an eclipse event inscribed. This correction on the Saros eclipse events cell numbering, reveal a new, very important observation: The preserved lunar eclipse event on the middle of cell-113 (initial cell-114) corresponds to the start of the second half of the Saros Cycle, i.e. a new Sar period starts from the middle of the cell. This period was first discussed by Ahnert, (1965) and the name Sar was suggested by Meeus, (1965). After one Sar period of 9.015$y$/19 eclipse seasons (Neugebauer, 1975; Oppolzer, 1889), the lunar phase presents a $\pi$-phase relative to the beginning of Saros (223/2 =111+0.5 synodic months= 111+one fortnight), on the same time the Node position presents a $2\pi$-phase (242/2 =121+0 draconic months) and also the lunar distance *Perigee/Apogee* presents a $\pi$-phase (239/2 = 119+0.5 anomalistic months).

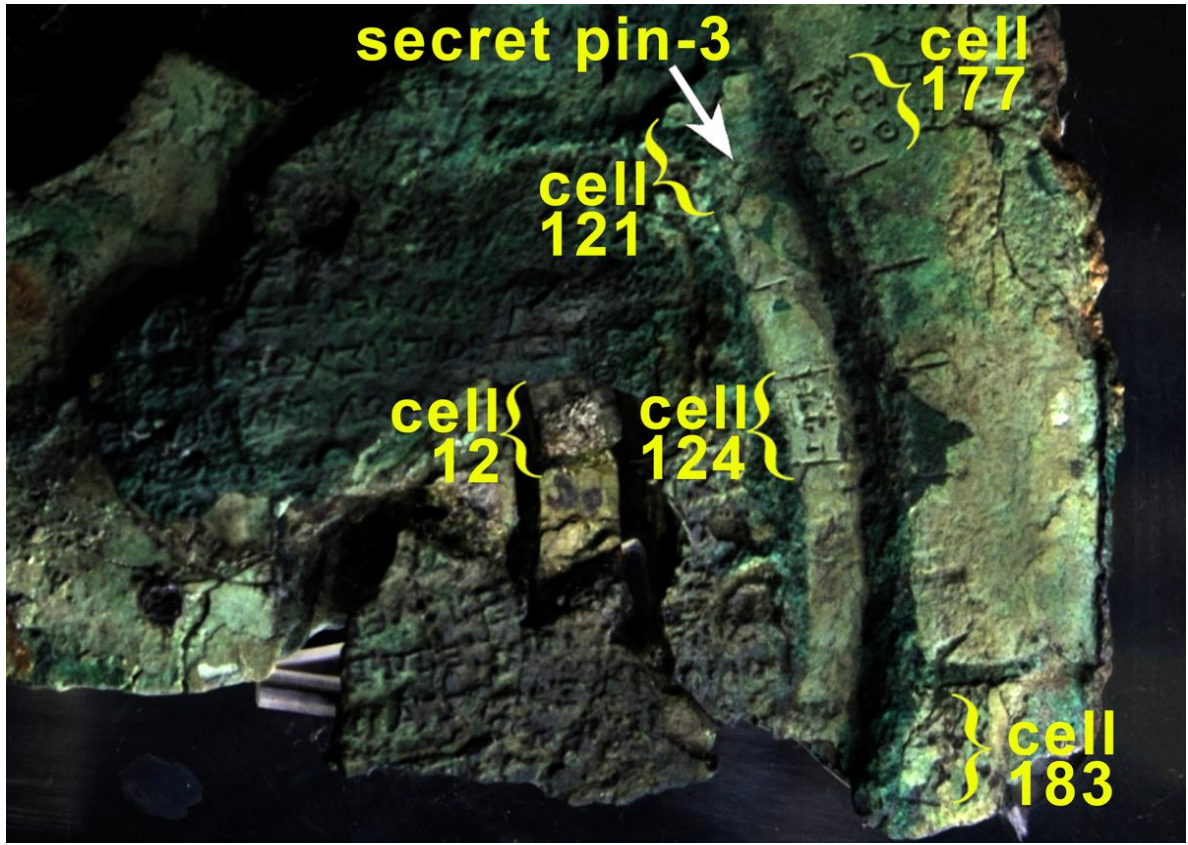

*Figure 11. The new numbering of the Saros cells with the preserved characteristic eclipse events is presented. AMRP PTM photograph processed by the authors.*

The calculation of 111.5 synodic =119.5 anomalistic =121 draconic months is very simple in mathematics, it is therefore quite sure that the ancient manufacturer could calculate it and use it. This equality presents a particular Symmetry for the corresponding eclipse events after one Sar period: Sun/Moon, Moon/Sun, Apogee/Perigee, Perigee/Apogee and Ascending (Descending) Node/Ascending (Descending) Node.

When a solar eclipse occurs in the northern/(southern) hemisphere, after a period of one Sar, the Moon crosses the northern/(southern) part of the Earth's umbral cone, thus leading to a lunar eclipse. Moreover, following a long duration total solar eclipse (i.e. Moon on or close to Perigee), a long duration total lunar eclipse will occur after one Sar (Full Moon on or close to Apogee). E.g. the Total Solar Eclipse of July, 22$^{nd}$ 2009, the longest in the 21$^{st}$ century in duration of 6m 39s during totality (https://eclipse.gsfc.nasa.gov/SEmono/TSE2009/TSE2009.html), as the Moon was well positioned at Perigee (http://www.fourmilab.ch/earthview/pa-calc.html), was followed by the Total Lunar eclipse of 27$^{th}$ July 2018, which was the longest total lunar eclipse of the 21$^{st}$ century in duration of totality 102m 57s, (https://eclipse.gsfc.nasa.gov/LEplot/LEplot2001/LE2018Jul27T.pdf), as the Moon was also well positioned at Apogee. The time span between these two eclipses is one Sar period. One Sar after the 2018 Lunar eclipse (one Saros cycle after the TSE of 2009), another long duration Total Solar Eclipse (6m 22sec) will take place on 2$^{nd}$ August, 2027, visible in Egypt and the Middle East, also in a parts of Atlantic and Indian Oceans Fig. 12 (http://xjubier.free.fr/en/site_pages/solar_eclipses/TSE_2027_GoogleMapFull.html). Obviously, the Earth-Moon distance (Apogee/Perigee) also determines the type of solar eclipse (total or annular) and affects its duration.

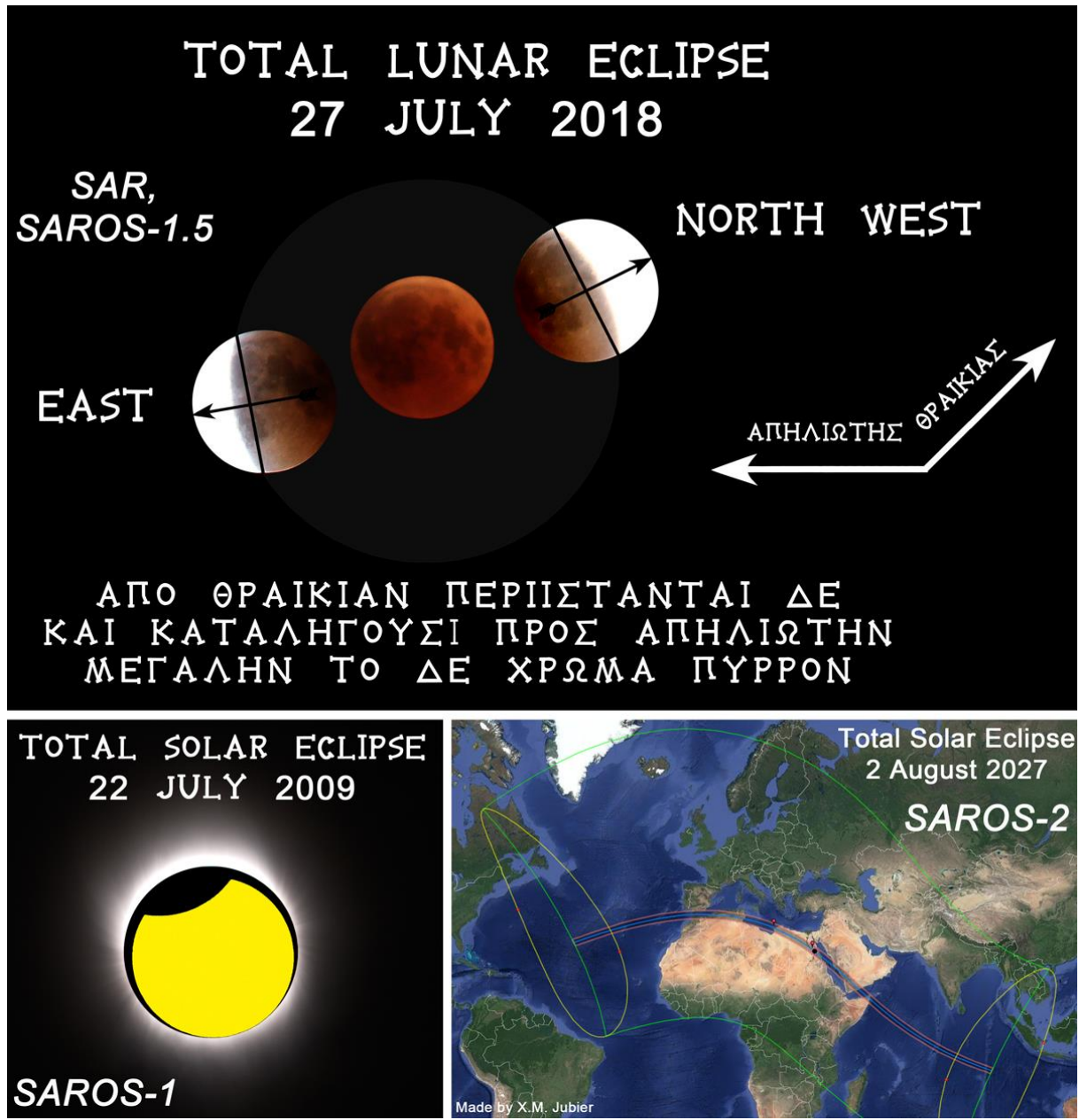

*Figure 12. Top, the Total Lunar Eclipse of July 27, 2018, observed from Thessaloniki, N. Greece, captured via a telescope. Three images were selected in order to artificially depict the Earth's shadow. Images by the first author. The path of the moon through the shadow is very similar to the Back Plate Inscription, "From Thrakias, and they veer about and end up towards Apêliôtês. Large(?). The color fiery red" (Anastasiou et al., 2016, page 163). The word "Large" is very probably referred to the event's duration. Bottom left, the Total Solar Eclipse of July 22, 2009, observed from Tian Huan Ping, Shanghai, China, captured via a telescope. The inner solar corona and the polar plumes are visible (Plutarch 1957, §19, p.118, also describes a total solar eclipse). A photograph from the partial phase (before totality), was digitally added on the Lunar Disc, in order to depict the diameter difference of the two celestial bodies. Images by the first author. Bottom right, the shadow path of the forthcoming very long duration Total Solar Eclipse of August 2, 2027, one Sar after the longest in 21th century, duration Lunar eclipse of 2018 and one Saros after the longest in 21th century, duration Total Solar eclipse of 2009 (http://xjubier.free.fr/en/site_pages/solar_eclipses/TSE_2027_GoogleMapFull.html).*

The variable distance of the moon was included by the ancient manufacturer's invention, constructing the *pin&slot* on the Antikythera Mechanism gearing (Freeth et al., 2006). The *pin&slot* reproduces the variable velocity of the Moon on the sky, which was known by the ancient Greeks, as Geminus 1880, 2002 describes. We strongly believe that the Saros spiral eclipse events, were calculated by the ancient manufacturer using the phase correlation of the lunar disc pointer aiming to the Golden Sphere-Sun (or in opposite position) and the proposed by the authors, new idea for the necessity of the existence of the Draconic gearing and pointer.

Only one eclipse event on cell-113 is engraved, a lunar eclipse (cell-114 for Freeth et al., 2008; Carman and Evans, 2014; Anastasiou et al., 2016; Freeth, 2019; Iversen and Jones, 2019; Jones, 2020). Therefore, one Sar period before the eclipsed Full moon, the New Moon would be located at about the same Node position as is the Full Moon of cell-113. This means that on cell-1/Saros spiral beginning, a solar eclipse (at least), should be engraved.

This observation leads to the conclusion that the Saros cycle of the Antikythera Mechanism dial starts on a synodic month/(date) in which a solar eclipse will occur. A solar eclipse occurred when the New moon is located on or close to a Node and therefore, after six synodic months from the Saros beginning (1+6), a lunar eclipse (or lunar and solar eclipse) should be (probably) occurred, which also agrees with the partially preserved events of cell-7 (initial cell-8) (Fig. 13).

Six pairs of cells with eclipse events in a time span of one Sar period are preserved, presenting "inversed" eclipse events (see the comments of Table 3 and Table 4).

*Table 3. The Saros spiral preserved eclipse events on the Fragments A2, F and E, and their corresponding event index letters, according to Freeth et al., 2008; Freeth, 2014; Anastasiou et al., 2016; Freeth, 2019; Iversen and Jones, 2019; Jones, 2020, is presented on 2nd on 3rd column. The fourth column presents the initial cell numbering of the eclipse possibility events (calculated before the present work, second/third column). The fifth column lists the corrected cell numbering after the positional apokatastasis of the Saros spiral.*

| Saros spiral turn | Event index letter | Eclipse possibility events of the Saros spiral cells (initial cell-number before the Saros spiral Apokatastasis) *Iversen and Jones 2019* | Initial cell numbering before the Saros spiral parts Apokatastasis (*Freeth et al., 2008; Freeth, 2014; Anastasiou et al., 2016; Freeth, 2019; Iversen and Jones, 2019; Jones, 2020*) | New cell numbering, after the Saros spiral Apokatastasis (present work) | Comments |
|---|---|---|---|---|---|
| 1st | B | Moon, daytime 2nd hour, Sun, 1st hour | (cell-8) | Cell-7 | One Sar period before Cell-119 |
|  |  |  | *Secret pin-1* | *Cell-10* | **Pin position on cell-10.4** |
|  | Γ | Sun, 1st hour | (cell-13) | Cell-12 | One Sar period before Cell-124 |
|  | E | Moon, 6th hour | (cell-20) | Cell-19 |  |
|  | Z | Sun, 6th hour | (cell-25) | Cell-24 | One Sar period before Cell-136 |
|  | H | Moon, daytime, 7th hour | (cell-26) | Cell-25 |  |
| 2nd | [Θ] | Moon, [- -?], [hour] | (cell-61) | Cell-60 |  |
|  | Π | Moon, nighttime, 8th hour | (cell-67) | Cell-66 |  |
|  |  |  | *Secret pin-2* | *Cell-66* | **Pin position on Cell-66.00** |
|  | Ρ | Sun, n[ighttime], 2nd hour | (cell-72) | Cell-71 | One Sar period before Cell-183 |
|  | Σ | See §2.5 of Iversen and Jones, 2019 | (cell-73) | Cell-72 |  |
|  | Τ | Sun, 1st hour | (cell-78) | Cell-77 | One Sar period before Cell-189 |
|  | Υ | Moon, daytime, 10th hour | (cell-79) | Cell-78 |  |
| 3rd | Γ̄ | Moon, daytime, 12th hour | (cell-114) | **Cell-113 Middle of *Saros Cycle*** | **On (middle) of Cell-113 (Full moon), a new Sar period (half Saros cycle) begins** |
|  | Δ̄ | Sun, nighttime, 12th hour | (cell-119) | Cell-118 |  |
|  | Ē | Moon, daytime, 12th(?) hour | (cell-120) | Cell-119 | One Sar period after Cell-7 |
|  |  |  | *Secret pin-3* | *Cell-121* | **Pin position on Cell-121.8** |
|  | Z̄ | Moon, daytime, 9th hour Sun, 3rd hour | (cell-125) | Cell-124 | One Sar period after Cell-12 |
| | Η̄ | Moon, 2nd hour, Sun, nighttime, 9th hour | *(cell-131)* | Cell-130 | |
| | Θ̄ | Moon, daytime 5th hour Sun, 12th hour | *(cell-137)* | Cell-136 | One Sar period after Cell-24 |
| 4th | Π̄ | Moon, 6th hour, Sun, 12th hour | *(cell-172)* | Cell-171 | |
| | Ρ̄ | Moon, 9th hour, Sun, 9th hour | *(cell-178)* | Cell-177 | |
| | | | *Secret pin-4* | *Cell-177* | **Pin position on Cell-177.5** |
| | Σ̄ | Moon, daytime 4th hour Sun, 1st hour | *(cell-184)* | Cell-183 | One Sar period after Cell-71 |
| | Τ̄ | Moon, daytime, 9th hour | *(cell-190)* | Cell-189 | One Sar period after Cell-77 |

Geminus, 1880 (also 2002; Evans and Berggren, 2006; Jones, 2017), refer that the Full moon occurring mid-month (15th day), named Διχόμηνις (Dichominis, Chap. VIII-"*About months*" and XI-"*About the Lunar eclipses*"), and the New moon occurring on the last day of month, named Τριακάς (Triakas, on 29th or 30th day, Chap. VIII, Chap. IX-"*About the light of the Moon*", and X "*About the Solar eclipses*"). Therefore, during the solar eclipse presented on cell-1, the Lunar disk pointer aims at the Golden sphere-Sun (New Moon - last day of month), and the Saros pointer must be placed at the end of cell-1, just before the boundary line of cell-2. Right after this position, a new month (Saros cell-2) begins on the Antikythera Mechanism dial-time measurements.

*Table 4. List of six pairs of preserved eclipse events. The (underline marked) cells on each pair have a time difference of one Sar and therefore the events are inversed.*

| **Sar period** | **Cell** | **Eclipse event** |
|---|---|---|
| Sar-I | 1 | (Sun) |
| Sar-II | 113 | Moon, daytime, 12th hour |
| Sar-I | 7 | Moon, daytime 2nd hour, Sun, 1st hour |
| Sar-II | 119 | Moon, daytime, 12th(?) hour |
| Sar-I | 12 | Sun, 1st hour |
| Sar-II | 124 | Moon, daytime, 9th hour Sun, 3rd hour |
| Sar-I | 24 | Sun, 6th hour |
| Sar-II | 136 | Moon, daytime 5th hour Sun, 12th hour |
| Sar-I | 71 | Sun, n[ighttime], 2nd hour |
| Sar-II | 183 | Moon, daytime 4th hour Sun, 1st hour |
| Sar-I | 77 | Sun, 1st hour |
| Sar-II | 189 | Moon, daytime, 9th hour |

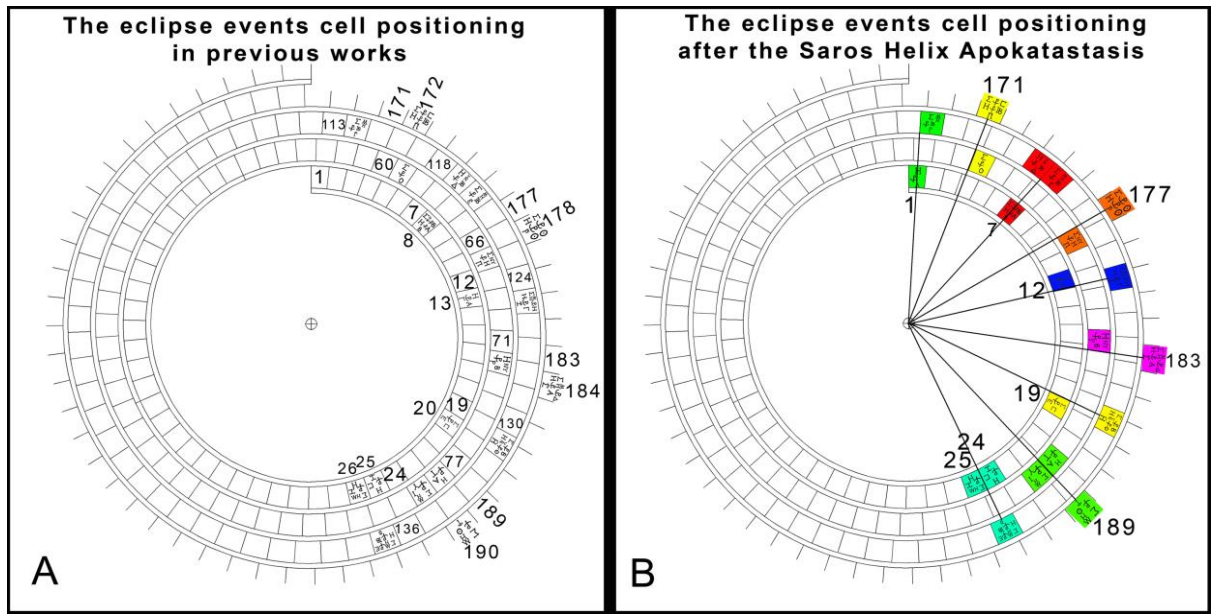

*Figure 13. A) The Saros spiral preserved eclipse events (Freeth et al., 2008; Freeth, 2014; Carman and Evans, 2014; Anastasiou 2016; Iversen and Jones, 2019), located on cell positions, according to the cell numbering of previous works. B) The Saros Helix eclipse events located on the new cell positions (corrected by minus 1), according to the revised cell numbering of present work. The lost eclipse event A, a solar eclipse (or both a lunar and solar eclipse) is located on cell-1 (one Sar period before the preserved cell 113, green color cells). The time span between the colored cells crossed by a ray line corresponds to a Sar period. Saros Helix starts on a month in which a solar eclipse event will occur at the last day of month (New moon phase). Images by the authors.*

### 5.2 An ancient Greek naming for the Saros spiral

The Saros cycle corresponds to the time duration between a repetition of a solar or lunar eclipse event with very similar characteristics (type and duration). This cycle was well-known, as is referred many times and was used by the ancient Assyrian, Babylonian, Egyptian and Greek astronomers. Thales from Miletus was probably used the Saros cycle, to predict the total solar eclipse of 28th May 585 BC, described by Herodotus. According to R.H. van Gent (https://eclipse.gsfc.nasa.gov/SEsaros/SEsaros.html) the word "Saros" was used by E. Halley in 1691, who read it in the 11th century Byzantine scholar lexicon Suidas (Bekkeri, 1854). Pliny the Elder (Naturalis Historia II.10,56) refers to the 223-month eclipse cycle without naming this period.

Ptolemy refers to the cycle of 223 synodic months (equal to 239 anomalistic cycles and 242 draconic cycles), that ancient astronomers referred to as *ΠΕΡΙΟΔΙΚΟΣ ΚΥΚΛΟΣ* or *ΠΕΡΙΟΔΙΚΟΣ ΧΡΟΝΟΣ* (Periodicos cycle or time). He also mentions that it was Hipparchus who re-measured and corrected the duration of Periodicos cycle to 4267 synodic months, equal to 4573 anomalistic cycles, 4612 minus 7.5° (4611 plus 352.5°) draconic cycles.

One of the Back cover plate preserved inscriptions of the Antikythera Mechanism (Bitsakis and Jones, 2016b, also phone communication with Prof. X. Moussas), is the phrase *ΤΗ ΕΛΙΚΙ ΤΜΗΜΑΤΑ ΣΛΕ* (in the entire spiral 235 sectors). This is directly correlated to the Metonic spiral, which was divided in 235 cells, representing the 235 synodic months of Enneakaidekaeteris (19th year cycle, Freeth et al., 2006; Freeth et al., 2008; Anastasiou et al., 2016). The ancient manufacturer, when referring to the word "*spiral*", he uses the Greek word *ΕΛΙΞ, ΕΛΙΚΑ* (Helix), instead of the word *ΣΠΕΙΡΑ* (Spiral). The word "*Helix*" in present day usually refers to a 3D geometrical curve, resembling a mechanical spring or screw, whereas the word "*spiral*" indicates a 2D geometrical curve, with its radius increasing relative to the epicenter angle. This mismatch probably started during the translation of Archimedes work "*ΠΕΡΙ ΕΛΙΚΩΝ*" (*About Helices*), in the 13th century, which was mistakenly translated into Latin as "*DE LINEIS SPIRALIBUS*" (*About spirals*) (Heiberg 1881, Israel 2015).

By combining the phrase *ΠΕΡΙΟΔΙΚΟΣ ΧΡΟΝΟΣ* (for Saros cycle) and the word *ΕΛΙΞ* (for Mechanism Spiral), a possible phrase for the Saros Spiral could be *Η ΤΟΥ ΠΕΡΙΟΔΙΚΟΥ ΚΥΚΛΟΥ ΕΛΙΞ* (Helix of Periodicos Cycle) or *Η ΤΟΥ ΠΕΡΙΟΔΙΚΟΥ ΧΡΟΝΟΥ ΕΛΙΞ* (Helix of Periodicos time). An alternative name could also be *Η ΕΛΙΞ ΤΩΝ ΕΚΛΕΙΨΕΩΝ* (Helix of the eclipses).

In the same manner, a respective name for the Metonic spiral could be *Η ΤΟΥ ΜΕΤΩΝΙΚΟΥ ΚΥΚΛΟΥ ΕΛΙΞ* or *Η ΤΟΥ ΜΕΤΩΝΟΣ ΕΝΝΙΑΥΤΟΥ ΕΛΙΞ* (Helix of Metonic cycle) or *Η ΕΝΝΕΑΚΑΙΔΕΚΑΕΤΗΡΙΔΟΣ ΕΛΙΞ* (Helix of Eneakedekaeteris).

## 6. CONCLUSION

The idea of Symmetry was extensively used in Ancient Greece, during the Proto-Geometric (Dark Ages 1050-900BC), Geometric (900-700BC), Archaic (700-500BC), Classical (500-323BC) (Horup, 2000; Coldstream, 1977; Roes, 1933; Schweitzer, 1971; Wide, 1899) and continued in the Hellenistic period, supported by Mathematics and Geometry. This idea is still in use today, as symmetry in mechanics, engineering, aerospace (Zipfel, 1975) etc., and it is very crucial for the trouble-free operation of a mechanical system, avoiding a large number of mechanical problems such as missing balance, rapid wear, poor stability, downside of efficiency etc.

The symmetry of the Antikythera Mechanism design is a very useful "tool" in order to have an opinion regarding the outer design of the Mechanism. It is also probable that, although the interior of the Mechanism was not directly visible as it was closed in a wooden box, the manufacturer tried to design it with symmetry.

This principle of Symmetry leads to the recalculation of the Saros axis position and the cell-number of the eclipse events, which the ancient manufacturer engraved in specific cells. This new numbering results to the conclusion that a solar eclipse event (at least) was engraved in cell-1, taking into account the half Saros period (Sar) and the preserved cells. Therefore, the Saros Helix of the Antikythera Mechanism, starts (cell-1) on a month in which a solar eclipse event will be occurred on 29th or 30th date (New moon phase). This observation introduces a new cycle of recalculations and discussions regarding the Epoch-Initial Date of the Antikythera Mechanism.

The argument that the Saros starts with a month which includes a solar eclipse and the second half Saros starts with a lunar eclipse, emphasizes - according to the authors' opinion - the attractive *Idea of Symmetry*, even on the engraved eclipse events on the Saros Helix.

**ACKNOWLEDGEMENTS**

We are very grateful to the AMRP for the license and permission to use the X-ray Data courtesy X-Tek Systems (now owned by Nikon Metrology), 2005 via equipment loaned by X-Tek System, Ltd. (now owned by Nikon Metrology) and also PTM data courtesy Hewlett-Packard, 2005. Thanks are due to the National Archaeological Museum of Athens, Greece for permission to photograph and study the Antikythera Mechanism fragments. We would also like to thank Prof. T. Economou of Fermi Institute-University of Chicago, USA, for his suggestions. First and second authors are member of a solar-eclipse research team of totality hunters headed by Prof. Jay M. Pasachoff. The team has travelled to China, Shanghai, Tian Huan Ping in 2009, in order to make white-light and spectroscopic observations of the solar corona, during the longest duration of totality of the 21th century. We thank the two anonymous referees for constructive comments.